\documentclass[conference]{IEEEtran}

\usepackage[utf8]{inputenc}
\usepackage{multirow}
\usepackage[inline]{enumitem}
\usepackage{xcolor}
\usepackage{booktabs}
\usepackage[pdfauthor={Maximilian Bachl, Joachim Fabini, Tanja Zseby}]{hyperref}
\usepackage{amsmath}
\usepackage{amssymb}
\usepackage{graphicx}
\usepackage[numbers]{natbib}
\usepackage{subfig}
\usepackage{tikz}
\usepackage{float}
\usepackage[OT1]{fontenc}

\usepackage[nomain, toc, acronym]{glossaries}
\glsdisablehyper

\begin{document}

\title{A flow-based IDS using Machine Learning in eBPF}

\author{\IEEEauthorblockN{Maximilian Bachl, Joachim Fabini, Tanja Zseby}
\IEEEauthorblockA{Technische Universität Wien\\
firstname.lastname@tuwien.ac.at}}





\maketitle%


\newacronym{ml}{ML}{Machine Learning}
\newacronym{dl}{DL}{Deep Learning}
\newacronym{ids}{IDS}{Intrusion Detection System}
\newacronym{rnn}{RNN}{Recurrent Neural Network}
\newacronym{dos}{DoS}{Denial-of-Service}
\newacronym{iat}{IAT}{Interarrival time}
\newacronym{nn}{NN}{neural network}
\newacronym{dt}{DT}{decision tree}
\newacronym{rf}{RF}{random forest}

\begin{abstract}
eBPF is a new technology which allows dynamically loading pieces of code into the Linux kernel. It can greatly speed up networking since it enables the kernel to process certain packets without the involvement of a userspace program. So far eBPF has been used for simple packet filtering applications such as firewalls or Denial of Service protection. We show that it is possible to develop a flow-based network intrusion detection system based on machine learning entirely in eBPF. Our solution uses a decision tree and decides for each packet whether it is malicious or not, considering the entire previous context of the network flow. We achieve a performance increase of over 20\% compared to the same solution implemented as a userspace program. 
\end{abstract}

\maketitle

\section{Introduction}

\subsection{eBPF}

eBPF is a technology which makes the Linux kernel programmable by enabling the injection of pieces of code at many locations of the kernel code. eBPF can be dynamically injected during runtime and is verified to make sure that it cannot crash and cannot get caught in infinite loops. However, this verification is only possible for programs that are not turing-complete. Thus eBPF programs cannot contain features such as loops of arbitrary length but instead loops must always have a maximum number of iterations. Also, backward jumps in the code are generally not allowed. This means eBPF can only be used to implement algorithms which do not require turing-completeness. eBPF programs are usually written in C and are first compiled to eBPF bytecode. Upon injection into the kernel, this eBPF bytecode is verified and dynamically compiled to native code. 

eBPF is especially suitable for packet processing: When a packet arrives at a network interface, certain actions can be performed such as dropping the packet. This is useful for programs such as firewalls \cite{risso_towards_2018} or \textit{tcpdump}, which records packets according to certain filters. For example, if only packets coming from port 80 should be recorded, tcpdump will compile an eBPF program which encodes this and will load it into the kernel. The kernel will then drop all packets which don't match the filter and only pass the correct ones to tcpdump. The alternative would be that tcpdump receives every packet and filters them itself. The drawback of this is that then each packet has to be passed from the kernel to tcpdump, which involves copying the whole packet in memory and also other computation steps. Thus, passing packets between the kernel and programs should be avoided if possible because of performance reasons. eBPF allows one to do that. 

Because eBPF bytecode is compiled to native code, it should generally be as fast as any other code in the kernel. A drawback, however, is that because eBPF is verified, it can only use certain data structures. For example, an eBPF program cannot use normal C arrays since they allow out-of-bounds accesses. For example, in an array of length 10, C would allow accessing the 15th element even though the array only has 10 elements. Thus, eBPF programs make use of special data structures which are safe. However, this can potentially be a performance penalty since checking the bounds of an array each time it is accessed requires extra work to be done by the CPU. 

One alternative to using eBPF is using kernel modules. However, a drawback of kernel modules is that they usually cannot be verified for stability and that they have to be compiled for a specific kernel version. Moreover, developing kernel modules is not straightforward and often it is not possible to extend certain functionality in the kernel with a kernel module without changing the kernel itself. Changing the whole kernel itself makes it necessary to recompile the kernel, which is cumbersome. 

\subsection{eBPF for a \gls{ml}-based \gls{ids}}

Some researchers have already investigated eBPF based \glspl{ids} or thought about using \gls{ml} in eBPF. \cite{ben-yair_ai_2019} propose to use AI for detecting performance anomalies with eBPF. However, they only propose a concept and do not provide an evaluation of the benefit of their approach. \cite{demoulin_detecting_2019}, \cite{van_wieren_signature-based_2019} and also \cite{choe_ebpfxdp_2020} use eBPF to develop solutions against Denial-of-Service attacks. They do not use \gls{ml}. 

One challenge of using eBPF for \gls{ml} is that it is not turing-complete. However, \gls{ml} algorithms such as \glspl{dt} or \glspl{nn} do not require loops in the implementation and thus don't require turing-completeness and can thus be implemented in eBPF. We decide to use \glspl{dt} since tree-based methods are a simple and effective \gls{ml} method for \glspl{ids} \cite{iglesias_ntarc_2020}. As mentioned in the previous sections, eBPF data structures need some additional processing compared to the classic ones built into C. A question we want to answer in this research is thus the following: Is eBPF faster than a solution implemented as a normal program in userspace? The userspace program has the disadvantage that all packets have to be passed between the kernel and the program which is slow. The eBPF program has the disadvantage that it makes use of potentially slower data structures. Thus, it is interesting to understand whether in practice eBPF can be faster even for complex programs which make use of data structures frequently. 

\section{Evaluation}

We envision an approach which keeps track of each network flow and analyzes each packet in the context of the previous packets of the flow. For example, certain attacks could be detected only when the fourth packet of the network flow containing the attack arrives. eBPF has built-in hash tables which allow storing information for each flow, which we use for this purpose. As the key we use the classic \textit{five tuple} of protocol type, source and destination IP and source and destination port. 

As features we use the source and destination port, the protocol identifier (UDP, TCP, ICMP etc.), the packet length, the time since the last packet of the flow and the direction of the packet (from sender to receiver or vice versa) because these features have proven useful for network traffic analysis \cite{iglesias_ntarc_2020}. Additionally, we include the mean of the packet size, the time since the last packet and the direction for all packets received in the flow so far. Furthermore, also the mean absolute deviation is computed for these three features, since the standard deviation cannot be computed due to the absence of advanced arithmetic operations in eBPF such as the square root operation.  A problem is that eBPF doesn't allow floating point operations but only integers. We thus implement the decision tree using fixed point arithmetic using 64 bit signed integers. We use 16 bits for the part after the fixed point.

We use the popular CIC-IDS-2017 dataset \cite{sharafaldin_toward_2018}. We train the decision tree using \textit{scikit-learn} with a maximum depth of 10 and a maximum number of leaves of 1000 using a train/test split of 2:1, which achieves an accuracy of 99\% on the testing dataset after training. This accuracy is comparable to the accuracy achieved in related work \cite{hartl_explainability_2020}. 

To enable reproducibility and to encourage further experimentation by other researchers, we make all the source code and other materials of this work publicly available \footnote{\url{https://github.com/CN-TU/machine-learning-in-ebpf}}.

We implement the same \gls{ids} in userspace and also in eBPF and use the previously trained \gls{dt}. The code is completely equivalent except that data structures are different because, as mentioned previously, many normal data structures are not usable in eBPF. Also, some data structures that eBPF has, such as hash maps, are not available in a normal C userspace program by default and thus we use a simple hash map implementation from the Linux kernel for the userspace version. 

We emulate a network using Linux network name spaces. For this, we have one server and one client, connected via a switch and configure the links to have no delay in addition to the delay caused by the Linux kernel for forwarding the packets. Furthermore, we set the maximum link speed to be unlimited. The server and the client are connected using \textit{iPerf}. Since the link speed is unlimited, iPerf will send as fast as possible. The maximum speed is only limited by how fast the computer can process the packets. The \gls{ids} is deployed by opening a raw socket on the network interface of the server. All packets passing through the server thus pass through the raw socket and are processed by the \gls{ids}. The \gls{ids} is implemented either as a classic userspace program or by using eBPF and run after one another (not at the same time). Both implementations are run for 10\,s. Instead of deploying the \gls{ids} on an end host, it can also be deployed on a router or switch, which runs a recent Linux version and thus can run eBPF. 

\begin{table}[h]
\caption{Maximum number of packets each implementation can process (mean and standard deviation). Results averaged over 10 runs each.} \label{tab:comparison}
\centering
\begin{tabular}{rrrrr} \toprule
& \multicolumn{2}{c}{Userspace} & \multicolumn{2}{c}{eBPF} \\ 
& Mean & SD & Mean & SD \\\midrule
packets/s & 125\,420 & 2627 & 152\,274 & 1201 \\
\bottomrule
\end{tabular}
\end{table}

\autoref{tab:comparison} shows that the userspace implementation inspects fewer packets per second (125420) than the eBPF implementation (152274). The eBPF implementation is thus more than 20\% faster than the userspace one. 

\section{Discussion}

An interesting consideration is that for complex \gls{ml} models, at some point the overhead from using eBPF's special data structures becomes larger than the benefit that eBPF confers. For example, interesting future work would be to test whether \glspl{rf} or deep \glspl{nn} can also achieve a performance advantage when implemented in eBPF. In this work, however, we could show that for a simple decision tree model, eBPF provides a significant performance advantage. 

\renewcommand*{\bibfont}{\small}
\bibliographystyle{ieeetr}
\bibliography{ml_in_ebpf}

\end{document}